\documentstyle[amssymb,aps,prb,multicol]{revtex}

\begin{document}
\draft
\title{Nonequilibrium Phase Transitions of Vortex Matter in Three-Dimensional
Layered Superconductors}
\author{Qing-Hu Chen$^{1,2}$ and Xiao Hu$^{1}$}
\address{$^{1}$Computational Materials Science Center, National Institute for
Materials Science, Tsukuba 305-0047, Japan\\
$^{2}$Department of Physics, Zhejiang University, Hangzhou 310027, P. R.China}
\date{\today}
\maketitle

\begin{abstract}
Large-scale simulations on three-dimensional (3D) frustrated anisotropic XY
model have been performed to study the nonequilibrium phase transitions of
vortex matter in weak random pinning potential in layered superconductors.
The first-order phase transition from the moving Bragg glass to the moving
smectic is clarified, based on thermodynamic quantities. A washboard noise
is observed in the moving Bragg glass in 3D simulations for the first time.
It is found that the activation of the vortex loops play the dominant role
in the dynamical melting at high drive.
\end{abstract}


\begin{multicols}{2}

Dynamical properties of current driven vortex matter \cite{Blatter,NS,GT}
interacting with random pinning potentials in type II superconductors have
attracted considerable attention both experimentally \cite
{BH,March,Par,tm,mae} and theoretically. \cite{KV,GL,BMR,SV} A better
understanding of various nonequilbrium phases and phase transitions is
essential for explaining the nonlinear current-voltage (I-V) characteristics
observed in experiments conducting on samples in an external magnetic field, 
\cite{BH,BSF} which is known as the first evidence of the first-order vortex
lattice melting. In addition, this problem is closely related to an
important class of phenomena in condensed matter physics, such as dynamics
of sliding charge-density waves (CDW) in quasi-one-dimensional conductors,\cite{CDW}
Wigner crystals in a two-dimensional (2D) electron gas, \cite{wigner} \ as
well as driven interface in random media.\cite{interface}

Vortex matter shows the three regimes of creep, depinning, and flow in its
transport characteristics, depending on the drive. In the flow regime, the
periodicity in the direction transverse to motion leads to a novel phase:
moving Bragg glass (BrG), based on an elastic transverse equation of motion
proposed by Giamarchi and Le Doussal (GL).\cite{GL} As the driven force is
reduced, the effective pinning strength becomes larger. It has been argued
that moving vortex matter may decay first into a moving smectic \cite{GL,BMR}
and then into a moving liquid. It can be further driven into a creeping BrG
below the depinning threshold. Recently, these moving phases have been
observed both experimentally \cite{March,Par} and in numerical simulations. 
\cite{Moon,MRR,ORN,DOM,van,OZKG,KDOG}

A precise analytical description is very challenging, especially in the
transition regime, because it is quite difficult to deal with the
topological defects\cite{SV2} in the plastic flow. Even within the elastic
approach, an analytical study of driven vortex matter is hampered by dynamic
nonlinearities such as Kardar-Parisi-Zhang term \cite{KPZ} that governs the
vortex dynamics on large scale. The underlying mechanism of 
the dynamical melting still remains open question. In
recent years, several three-dimensional (3D) numerical simulations have been
performed.\cite{van,OZKG,KDOG} However, one common shortage is that the
moving BrG holds out to arbitrary high drives, which is obviously
contradictory to any real experiments.

In the present Letter, we report new results of nonequilibrium simulations
for vortex matter in anisotropic 3D systems with weak disorder. The
dynamical melting from the moving BrG to the moving smectic is found to be
first order. The washboard noise is observed in the voltage power spectra in
the moving BrG. Our results suggest that the thermally activated vortices
play a dominant role in the dynamical melting at high currents.

Vortex matter in type II layered superconductors can be described by the 3D
anisotropic XY model on a simple cubic lattice \cite{hu,HO,OT} 
\begin{equation}
H=-\sum_{\langle ij\rangle }J_{ij}\cos (\phi _{i}-\phi _{j}-A_{ij}),
\label{Hamil}
\end{equation}
where $\phi _{i}$ specifies the phase of the superconducting order parameter
on site $i$, \ $A_{ij}=(2\pi /\Phi _{0})\int_{i}^{j}{\bf A\cdot dl}$ with $
{\bf A}$ the magnetic vector potential of a field ${\bf B}=\nabla \times 
{\bf A}$ along the $z$ axis. The random pinning potential is introduced in
the coupling strength in the $xy$ plane $J_{ij}=J_{0}(1+p\epsilon _{ij})$,
where $\epsilon _{ij}$'s are independently Gaussian distributed with zero
mean and unit variance, $p$ represents the pinning strength.\cite{OT} The
coupling between the $xy$ planes is $J_{z}=J_{0}/\Gamma ^{2},$ ($\Gamma $ is
the anisotropy constant). This model is relevant to high-$T_{c}$
superconductors and artificially layered superconductors. For the data
presented below, we typically choose $p=0.04$ which models weak pinning
strength, $1/\Gamma ^{2}=1/40$, and the average number of vortex lines per
plaquette $f=l^{2}B/\Phi _{0}=1/20$, where $l$ is grid spacing in the ab
plane. Our system size is $L=40$ for all directions.

In order to study the transport properties, we incorporate the
Resistivity-Shunted-Junction dynamics in simulations. Realizing that the sum
of supercurrents into site $i$ can be expressed in terms of the derivative
of  Eq. (1)  with respect to $\phi _{i}$, the dynamical equations for the $
\phi $'s are readily derived by requiring the sum of currents into each site
to vanish 
\begin{equation}
{\frac{\sigma \hbar }{2e}}\sum_{j}(\dot{\phi _{i}}-\dot{\phi _{j}})=-{\frac{
\partial H }{\partial \phi _{i}}}+J_{{\rm ext},i}-\sum_{j}\eta _{ij},
\end{equation}
where $J_{{\rm ext},i}$ is the external current which vanishes except for
the boundary sites. The $\eta _{ij}$ is the thermal noise current with zero
mean and a correlator $\langle \eta _{ij}(t)\eta _{ij}(t^{\prime })\rangle
=2\sigma k_{B}T\delta (t-t^{\prime })$. In the following, the units are
taken of $2e=J_{0}=\hbar =\sigma =1$.

In the present simulation, a uniform external current $I_{x}$ along $x$
-direction is fed into the system. The fluctuating twist boundary condition 
\cite{MKO} is applied in the $xy$ plane to maintain the current, and the
periodic boundary condition is employed in the $z$ axis. In the $xy$ plane,
the supercurrent between sites $i$ and $j$ is now given by $\
J_{i\rightarrow j}^{(s)}=J_{ij}\sin (\theta _{i}-\theta _{j}-A_{ij}-{\bf r}
_{ij}\cdot {\bf \Delta }),$with$\;{\bf \Delta }=(\Delta _{x},\Delta _{y})$
the fluctuating twist variable and $\theta _{i}=\phi _{i}+{\bf r}_{i}\cdot 
{\bf \Delta }$. The new phase angle $\theta _{i}$ is periodic in both $x$-
and $y$-directions. Dynamics of ${\bf \Delta }$ can be then written as 
\begin{equation}
\dot{\Delta}_{\alpha }={\frac{1}{L^{3}}}\sum_{<ij>_{\alpha
}}[J_{i\rightarrow j}^{(s)}+\eta _{ij}]-I_{\alpha },  \label{delta-dot}
\end{equation}
where $\alpha =x,y$. The voltage drop is $V=-L{\dot{\Delta}_{x}}$.

The above equations can be solved efficiently by a pseudo-spectral algorithm 
\cite{chen} due to the periodicity of phase in all directions. The time
stepping is done using a second-order Runge-Kutta scheme with $\Delta t=0.05$%
. Our runs are typically $(4\thicksim 8)\times 10^{7}$ time steps and the
latter half time steps are for the measurements. For a given current, the
simulations are started from high temperatures with random initial phase
configurations, and vortex systems are gradually cooled down. The present
results are based on one realization of disorder. For these parameters, the
equilibrium BrG melts to a liquid at $T=0.254$\cite{OT} and the critical
current $I_{c}(T=0)$ is estimated to be around $0.095$. We calculate the
internal energy $e$ per flux line per layer, the helicity modulus along the $%
z$ axis $\Upsilon _{z}$, and the vortex structure factor in the $xy$ plane,
together with the density of dislocations $\rho _{d}$ and vortex-antivortex
pairs $\rho _{av}$ in the $xy$ plane. The helicity modulus describes the
superconducting phase coherence.

As shown in Fig. 1, evident jump of $e$ is observed around the melting
temperature $T_{m}$ for $I=0.5$, which clearly indicates a first-order
dynamical melting. The helicity modulus $\Upsilon _{c}$ is very sharply set
up precisely at $T_{m}$, consistent with the first-order nature. The vanish
of $\Upsilon _{z}$ above $T_{m}$ shows that the dynamical melting marks the
loss of superconducting phase coherence along the $z$ direction.

To characterize the spatial order of moving phases, we show the structure
factors in the vicinity of $T_{m}$ at various currents in Fig. 2. At high
currents $I=0.5$ and $0.3$($\gg I_{c}(0)$), moving BrG's with six
well-defined peaks are found just below $T_{m}$, which shows the
quasi-long-range translational correlations both perpendicular and parallel
to the flow direction, consistent with the predictions by GL.\cite{GL} By
mapping out the trajectories of the moving vortices, we found a  set of
periodic coupled elastic channels, corresponding to peaks in the $K_{y}=0$
axis. The peak in $K_{y}\neq 0$ is anisotropic. The half width along the
flow direction is considerably smaller than that transverse to the flow,
indicating that the positional correlation along the flow direction is much
stronger than that transverse to the flow, which provides a solid base for
the moving BrG theory, \cite{GL} where a long range order is assumed along
the flow direction.

One can see from Figs. 2 (a) and (b) that the moving BrG melts sharply into
a moving smectic just above $T_{m}$ where Bragg peaks only remain in $
K_{y}=0 $ axis, confirming the proposal by Balents et al. \cite{BMR} The
current dependent anisotropy of the smectic peak will be discussed later.
The moving smectic was unnoticed in a previous simulation \cite{DOM} using
isotropic $3D $ XY model for higher vortex density with smaller system size.
In addition, the orientation of vortex lattice was found to be not aligned
with the direction of the motion, \cite{DOM} inconsistent with the
characterization of the moving BrG.

Through extensive simulations in vortex flow regime, we universally find
that the appearance of a moving smectic is simultaneously accompanied by the
loss of the superconducting phase coherence along the $z$ direction ($
\Upsilon _{z}=0$). It demonstrates that the vortices flow incoherently in
different $xy$ planes in a moving smectic.

As the temperature further increases, the moving smectic continuously
evolves into a moving liquid with a ringlike pattern in the structure factor
(not shown here). We have not found any anomaly in $e$, demonstrating a
continuous phase transition. In low current regime, e.g. $I=0.05$ ($\ll
I_{c}(0)$), it is found in Fig. 2(c) that the (creeping) BrG directly melts
to a liquid without through an intermediate moving smectic, close to
equilibrium.

To capture insight of the underlying mechanism of the dynamical melting, we
display temperature dependence of the density of dislocations $\rho _{d}$ in
Fig. 3. Sharp jump of $\rho _{d}$ right at $T_{m}$ demonstrates that the
dynamical melting is mediated by the proliferation of dislocations,
analogous to the equilibrium melting\cite{HO}. Dislocations can be created
either by entanglement of the field-induced flux lines or by excitations of
small vortex loops, the latter could be more important in the presence of
external currents. For $I=0.3$, we find that vortex loops are only activated
far above $T_{m}$, which excludes the effect of vortex loops in the melting.
Anisotropy of the smectic peak at this current exhibited in Fig. 2(b) is
therefore quite similar to that in Ref. (19) without the mechanism to
generate vortex loops. However, at $I=0.5$, the thermally activated vortices
play an essential role in the dynamical melting. The density of thermally
excited vortices $\rho _{av}$ also shows a steep jump at $T_{m}$. The
smearing out of the smectic peak transverse to the flow at $I=0.5$ shown in
Fig. 2(a) is just originated from the nucleation of vortex loops. As the
current increases to $0.69$, ''thermally'' activated vortex-antivortex pairs
remain down to zero temperature. Above this current, the superconducting
coherence along the $z$ direction vanish at any temperature. In this sense,
we refer this current to a depairng critical current. It is just the
thermally activated vortex loops that induce dislocations, which in turn
destroy the superconducting moving BrG in very high currents. The
perturbation of these excited vortex loops has not been considered in the
analytical theories. \cite{GL,BMR,SV} In previous simulations
using Langvein dynamics of a fixed number of interacting particles, \cite
{Moon,MRR,ORN,van,OZKG,KDOG} the effect of thermally activated vortices are
absolutely excluded, so the moving BrG remains unreasonably in arbitrary
high currents.

Repeating such simulations at various currents, we compose a dynamical phase
diagram of vortex matter as a function of $I$ and $T$ in Fig. 4. The
continuous phase transition from the moving smectic to the moving liquid is
determined by criterion that the smectic peak height from the ring
background is less than $0.05$. $T_{m}(I)$ is always lower than $T_{m}(0)$
and shows a nonmonotonic characteristics. This naturally bring about the
possibility of the reordering of driven vortex matter with the increase of
drives predicted theoretically.

To show the temporal correlation in BrG phase, we calculate the power
spectra of voltage noise $s(\omega )=\left| \int V_{x}(t)e^{-i2\pi \omega
t}dt\right| ^{2}$. In Fig. 5(a), well-developed peaks appear at the
washboard frequency and harmonics for $I=0.3$ at $T=0$. The voltage noise
spectrum for $I=0.5$ at $T=0$ is also presented in the inset of Fig. 5(a)
for comparison. The voltage cross the sample is $V_{x}(t)\varpropto nv_{y}$,
where $n$ is the vortex density and $v_{y}$ is the vortex velocity in the $y$
direction. \cite{BS} The washboard frequency is given by $\omega _{0}=v_{y}/a
$, where $a$ is the vortex spacing. Due to the same vortex density, we
really find that the voltage ratio ($\thickapprox 2.01$) is exactly
consistent with that of two washboard frequencies. As the temperature
increases, as shown in Fig. 5(b), the peak at $\omega _{0}$ remains in all moving BrG
phases and suddenly disappears just above $T_{m}$. The
moving BrG phase is then characterized by the peak at $\omega _{0}$.
Interestingly, we find that the peak at $\omega =5\omega _{0}$
owing to the network discretization is gone in temperatures far below $T_{m}
$, which excludes the effect of the artificial lattice pinning in the phase
transition. In the moving smectic, the short range of the position
correlation along the flow direction totally destroys the temporal order of
the moving vortices. To the best of our knowledge, the washboard noise,
observed recently in both conventional\cite{tm} and high-$T_{c}$
superconductors, \cite{mae} has never been reported in numerical simulations
except in 2D at zero temperature.\cite{ORN} We believe this to be the first
observation of the washboard velocity modulation of driven vortex matter in
3D numerical simulations.

We acknowledge useful discussions with A. Maeda, Y. Nonomura, L. H. Tang, A.
Tanaka, and H. H. Wen. The present simulations were performed on the
Numerical Materials Simulator (SX-5) in NIMS, Japan. This study was
supported by the Ministry of Education, Culture, Sports, Science and
Technology, Japan, under the Priority Grant No. 14038240. One of us (QHC)
was partially supported by the JSPS invitation program and the NSFC under
Grant No. 10075039 \& 10274067.

Fig. 1. Temperature dependence of internal energy $e$ (cycles) and helicity
modulus along the $z$ axis $\Upsilon _{z}$ \ (squares) at $I=0.5$.

Fig. 2. The
vortex structure factors just above and below $T_{m}$ for (a) $T=0.5$, (b) $ 
T=0.3$, and (c) $T=0.05$.

Fig. 3.Temperature dependence of the density of
dislocations $\rho _{d}\;$and thermally activated vortex-antivortex pairs $ 
\rho _{av} $at various currents.

Fig. 4. Dynamical phase diagram in
temperature-current plane. Solid lines with open cycles: $1^{st}$ order
(dynamical) melting from moving BrG to smectic. Dashed lines: continuous
phase transition from the moving smectic to liquid. 

Fig. 5. Voltage noise power
spectra $s(\omega )$ for $I=0.3$ at (a) $T=0$, (b) $T=0.118 $, and for $I=0.5$
at $T=0$ in the inset of (a).

\end{multicols}

\end{document}